\documentstyle[12pt,psfig,draft]{nature-dbf}

\newcommand{\grb}{GRB\,050709}

\newcommand{\hete}{HETE}
\newcommand{\hetelong}{High-Energy Transient Explorer}
\newcommand{\hst}{HST}
\newcommand{\hstlong}{Hubble Space Telescope}
\newcommand{\Swift}{Swift}
\newcommand{\swift}{Swift}

\newcommand{\chandra}{Chandra}
\newcommand{\chandralong}{Chandra X-ray Observatory}

\def\simlt{\mathrel{\hbox{\rlap{\hbox{\lower4pt\hbox{$\sim$}}}\hbox{$<$}}}}
\def\simgt{\mathrel{\hbox{\rlap{\hbox{\lower4pt\hbox{$\sim$}}}\hbox{$>$}}}}

\def\arcsec{\hbox{$^{\prime\prime}$}}

\newcommand{\Msunyr}{\mbox{$M_\odot$ yr$^{-1}$}}

\newcommand{\xray}{\mbox{X-ray}}

\newcommand{\uJy}{\mbox{$\mu$Jy}}

\newcommand{\ergsec}{\mbox{erg s$^{-1}$}}
\newcommand{\ergcmsq}{\mbox{erg cm$^{-2}$}}
\newcommand{\ergcms}{\mbox{erg cm$^{-2}$ s$^{-1}$}}

\newcommand{\kmsecmpc}{\mbox{km s$^{-1}$ Mpc$^{-1}$}}

\newcommand{\percmsq}{cm$^{-2}$}
\newcommand{\percmcube}{cm$^{-3}$}

\begin{document}

\title{The afterglow of GRB\,050709 and the nature of the short-hard
       $\gamma$-ray bursts}

\author{%
   D.~B.~Fox\affiliation[1]
       {Division of Physics, Mathematics \&\ Astronomy,
       California Institute of Technology, Pasadena, CA\, 91125,
       USA\vspace{0.05in}}$^{,}$%
\affiliation[2]%
      {Department of
       Astronomy \& Astrophysics, 525 Davey Laboratory,
       Pennsylvania State University, University Park, PA\, 16802,
       USA\vspace{0.05in}}
D.~A.~Frail,\affiliation[3]%
     {National Radio Astronomy Observatory, P.O. Box O,
      Socorro, NM\, 87801, USA\vspace{0.05in}}
P.~A.~Price,\affiliation[4]%
       {University of Hawaii, Institute for Astronomy,
       2680 Woodlawn Drive, Honolulu, HI\, 96822, USA\vspace{0.05in}}
S.~R.~Kulkarni,\affiliationmark[1]
E.~Berger,\affiliation[5]%
     {Carnegie Observatories, 813 Santa Barbara Street,
     Pasadena, CA\, 91101, USA\vspace{0.05in}}
T.~Piran,\affiliationmark[1]$^{,}$\affiliation[6]%
        {Racah Institute for Physics, The Hebrew
        University, Jerusalem 91904, Israel\vspace{0.05in}}
A.~M.~Soderberg,\affiliationmark[1]
S.~B.~Cenko,\affiliationmark[1]
P.~B.~Cameron,\affiliationmark[1]
A.~Gal-Yam,\affiliationmark[1]
M.~M.~Kasliwal,\affiliationmark[1]
D.-S.~Moon,\affiliationmark[1]
F.~A.~Harrison,\affiliationmark[1]
E.~Nakar,\affiliationmark[1]
B.~P.~Schmidt,\affiliation[7]%
     {Research School of Astronomy and Astrophysics, The
      Australian National University, Weston Creek, ACT 2611,
      Australia\vspace{0.05in}},
B.~Penprase,\affiliation[8]%
   {Pomona College, 610 North College Avenue, Claremont,
   CA\, 91711, USA\vspace{0.05in}}
R.~A.~Chevalier,\affiliation[9]%
       {Department of Astronomy, University of Virginia,
        P.O. Box 3818, Charlottesville, VA\, 22903, USA\vspace{0.05in}}
P.~Kumar,\affiliation[10]%
       {Astronomy Department, University of Texas, Austin,
       TX\, 78731, USA\vspace{0.05in}}
K.~Roth \affiliation[11]%
       {Gemini Observatory, 670 North A`ohoku Place, Hilo,
       HI\, 97620, USA\vspace{0.05in}},
D.~Watson,\affiliation[12]%
       {Niels Bohr Institute, University of Copenhagen,
        Juliane Maries Vej 30, DK-2100 Copenhagen,
       Denmark\vspace{0.05in}}
B.~L.~Lee,\affiliation[13]%
       {Department of Astronomy \& Astrophysics,
       University of Toronto, Toronto, Ontario, M5S 3H8,
       Canada\vspace{0.05in}}
S.~Shectman,\affiliationmark[5]
M.~M.~Phillips,\affiliationmark[5]
M.~Roth,\affiliationmark[5]
P.~J.~McCarthy,\affiliationmark[5]
M.~Rauch,\affiliationmark[5]
L.~Cowie,\affiliationmark[4]
B.~A.~Peterson,\affiliationmark[7]
J.~Rich,\affiliationmark[7]
N.~Kawai,\affiliation[14]%
        {Department of Physics, Tokyo Institute of
         Technology, Ookayama 2-12-1, Meguro-ku, Tokyo 152-8551,
         Japan\vspace{0.05in}}
K.~Aoki,\affiliation[15]%
        {Subaru Telescope, National Astronomical
         Observatory of Japan, 650 N. A'ohoku Place, Hilo, Hawaii
        76720, USA\vspace{0.05in}}
G.~Kosugi,\affiliationmark[15]
T.~Totani,\affiliation[16]%
        {Department of Astronomy, School of Science, Kyoto University,
        Sakyo-ku, Kyoto 606-8502, Japan\vspace{0.05in}}
H.-S.~Park,\affiliation[17]%
       {Lawrence Livermore National Laboratory, 7000 East
       Avenue, Livermore, CA\, 94550, USA\vspace{0.05in}}
A.~MacFadyen\affiliation[18] %
       {Institute for Advanced Study, Princeton, NJ\, 08540,
       USA\vspace{0.05in}} 
\&\
K.~C.~Hurley\affiliation[19]%
       {Space Sciences Laboratory, University of
       California, Berkeley, CA\, 94720, USA\vspace{0.05in}}
}
\date{}{}
\headertitle{The afterglow of GRB\,050709}
\mainauthor{Fox~\textit{et al.}}


\summary{\small\bf%
The final chapter in the long-standing mystery of the gamma-ray bursts
(GRBs) centres on the origin of the short-hard class, suspected on
theoretical grounds to result from the coalescence of neutron star or
black hole binary systems. Numerous searches for the afterglows of
short-hard bursts have been made, galvanized by the revolution in our
understanding of long-duration GRBs that followed the discovery in
1997 of their broadband (\xray, optical, and radio) afterglow
emission. Here we present the discovery of the \xray\ afterglow of a
short-hard burst whose accurate position allows us to unambiguously
associate it with a star-forming galaxy at redshift $z=0.160$, and
whose optical lightcurve definitively excludes a supernova
association. Together with results from three other recent short-hard
bursts, this suggests that short-hard bursts release much less energy
than the long-duration GRBs.  Models requiring young stellar
populations, such as magnetars and collapsars, are ruled out, while
coalescing degenerate binaries remain the most promising progenitor
candidates.}

\maketitle


\section{Introduction}

High-energy transients remain at the frontier of astrophysics research
because they probe extreme physical regimes of matter, gravity, and
energy density. Soft $\gamma$-ray repeaters (SGRs) combine matter at
supra-nuclear densities with magnetic fields in excess of
$10^{15}$\,G, while long-duration GRBs, which probably herald the
birth of stellar-mass black holes, drive ultra-relativistic outflows
and power the brightest explosions in the Universe.  Progress in
understanding new classes of high-energy transient has typically
required a multiwavelength approach; in particular, the identification
of longer-wavelength counterparts enables precision localization and
detailed studies. It was the discovery of the slow-fading `afterglow'
emission of long-duration GRBs that enabled their sub-arcsecond
localization, the measurement of their redshifts, the identification
of their star-forming host galaxies, the quantification of their
energy scale, and ultimately, established their connection to the
deaths of massive stars (see ref \pcite{zm04} for a review).

The nature of the short-hard gamma-ray bursts (SHBs) has been an
outstanding mystery of high-energy astrophysics for more than
30~years.  The SHBs comprise about 30\% of the GRB population at the
Burst and Transient Source Experiment (BATSE) threshold and have
typical durations of 0.3\,s and peak energies of order 350\,keV, with
power-law tails extending to higher energies.\cite{kmf+93} Despite
extensive searches,\cite{hbc+02} no short-hard burst has yet been
sufficiently well-localized to ascertain its origins.  Historically,
$\gamma$-ray satellites were either not sensitive to SHBs or provided
positions that were too crude or too delayed to enable deep
searches. A significant breakthrough came when the \xray\ telescope
(XRT) on the recently-launched \swift\ satellite detected the
rapidly-fading afterglow of GRB\,050509B and localized it to a
circular region of radius 9.3 arcseconds. Within this \xray\
localization there are nearly 50 objects identified in \hstlong\
(\hst) images,\cite{kfc+05} the brightest of which by far is an
elliptical galaxy at $z=0.2248$ that has been proposed as the likely
host galaxy of this burst.\cite{bpp+05,gbb+05}

Even without any SHB distance determinations, the isotropic sky
distribution and non-euclidean brightness distribution of the SHBs
suggest a cosmological origin,\cite{kmf+93,schmidt01} fuelling
speculation that short-hard bursts are the result of the coalescence
of compact object (neutron star-neutron star or black hole-black hole)
binaries.\cite{elp+89} Theoretical estimates yield merger
rates\cite{gp05} that can easily accommodate the observed burst rate,
with engine lifetimes and energy releases roughly consistent with the
burst properties for a cosmological population.  Nonetheless, without
any detailed knowledge of their distances, energetics, and
environments, younger progenitor populations such as magnetars and
collapsars cannot be ruled out. If the coalescence model is correct,
the SHBs will be a primary source population for the Laser
Interferometer Gravitational Wave Observatory and other ground-based
gravitational wave detectors.  As such, the SHBs promise to provide a
crucial test-bed for theories of strong-field gravity, the nuclear
equation of state and the formation of black holes.


\section{Discovery of the X-ray afterglow}

\begin{figure}
\centerline{~\psfig{file=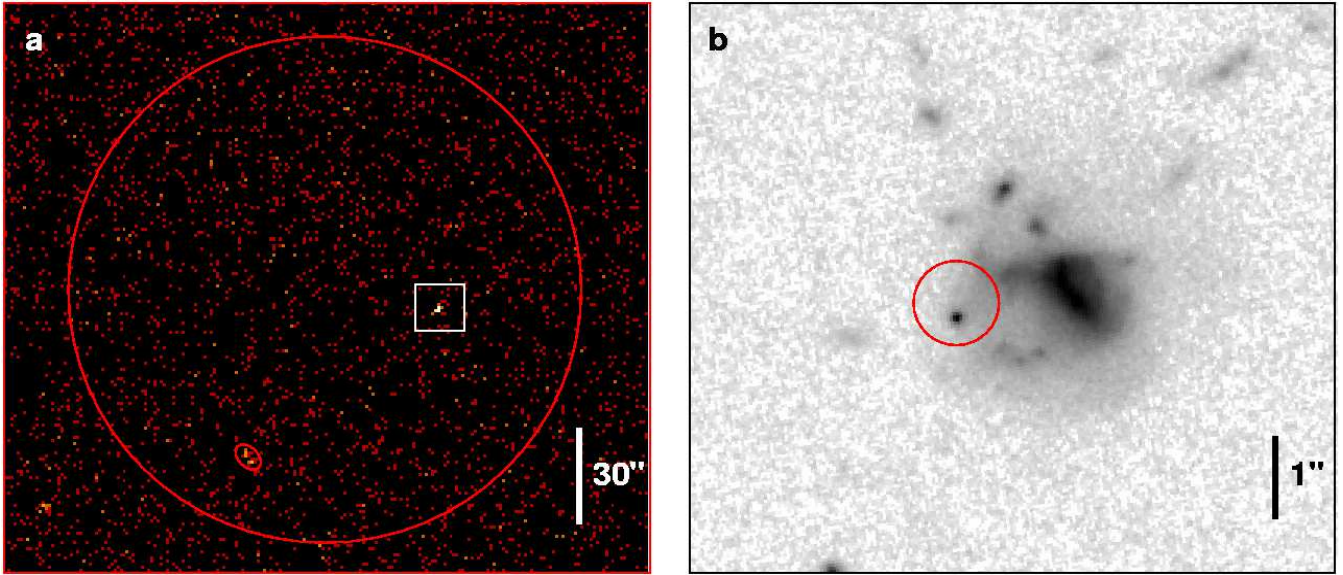,width=7.0in}~}
\caption[]{\bf HST and Chandra X-ray Observatory images of
    the afterglow and environs of GRB\,050709}
\label{fig:HST}
\end{figure}

Upon receiving notification of the localization\cite{vlr+05} of the
short-hard burst \grb\ by the \hetelong\ (\hete), we initiated
observations with the \chandralong\ as part of our approved program
for the SHBs, observing the 81-arcsec error circle with the Advanced
CCD Imaging Spectrometer (ACIS)\cite{gbf+03}.  A total of 38.4\,ks of
good data were obtained after excluding intervals of background
flaring activity, at a mean epoch of 2.52 days after the burst
(Table~\ref{tab:fluxes}).  We detected two sources in the \hete\ error
circle: a faint and resolved (or double) source and a bright point
source (see Figure~\ref{fig:HST}).  The faint source is well detected
at low energies (0.3--2.0\,keV band), has a flux of $3.0\times
10^{-15}$\,\ergcms\ (1--5 keV band) and coincides with a catalogued
radio source from the National Radio Astronomy Observatory Very Large
Array (VLA) \mbox{20-cm} Sky Survey\cite{ccg+98} (NVSS).

The bright point source has $49.5\pm 8.8\,$counts in the 0.3--8\,keV
band, corresponding to an \xray\ flux of $3.5\times
10^{-15}$\,\ergcms\ (1--5\,keV band) for the best-fit power-law
spectrum (photon index $\alpha=2.24\pm 0.35$, with column density
fixed to the expected Galactic hydrogen column density, $N_{\rm
H}=1.2\times 10^{20}$\,\percmsq).  The source is located at $\alpha =
\mbox{23:01:26.96}$, $\delta = -\mbox{38:58:39.5}$ (J2000).  This
position has been corrected by 0.4~arcsecond from the native \chandra\
astrometry using three X-ray sources coincident with stars in the
United States Naval Observatory (USNO) B1.0 catalog. We estimate the
90\% confidence radius is 0.5 arcsecond.

We proposed\cite{ffc+05} the brighter source as the \xray\ afterglow
of \grb. We also noted that it was offset by about one arcsecond from
a faint ($R\sim 20.5$\,mag) source visible in the Digitized Sky
Survey, plausibly its host galaxy. Making use of the \chandra\
position, we find a marginal detection of the \xray\ source in earlier
observations made by the \swift\ XRT (Table~\ref{tab:fluxes}).

We then executed an \mbox{18-ks} follow-up observation (mean epoch
16.0 days post-burst) with an identical observatory configuration.
These data showed that, at a 99.7\% confidence level, the \xray\
source had faded, roughly by a factor of two.  Inspecting the eleven
events within a 1.5-arcsec radius of the \xray\ afterglow position, we
found that nine events occurred within the first third of the
observation.  This is in contrast to the first epoch, which exhibits a
roughly uniform count rate over the observation. A Kolmogorov-Smirnov
(KS) test demonstrates that the second epoch arrival times are
inconsistent with a steady event rate at 99.9\% confidence.  We
therefore suggest that during the first $\sim$6\,ks of this
observation the source was in a ``flaring'' state, roughly an order of
magnitude brighter than during the remainder of the observation.

In Table~\ref{tab:fluxes} we give the mean epochs and \xray\ fluxes
for the flaring and quiescent portions of this observation.  Using the
quiescent flux, which represents only a marginal detection of emission
(90\% confidence), we find a temporal decay index in the \xray\ band
of $\alpha_{\rm X}\simlt -1$ for the interval from 2.5 to 16 days
after the burst.

The discovery of this flaring behavior suggests that even 16~days
after the burst, the afterglow is still subject to new energy inputs.
The sudden cessation of the flare represents a small fraction of the
time since the burst, indicating that the flaring must arise from a
source physically distinct from the fading afterglow. We suggest that
the flare arises from ongoing activity of the central engine, in
analogy to the bright \xray\ flares observed from several
long-duration \swift\ GRBs\cite{brf+05}.


\section{Optical afterglow and host galaxy}

In addition to our \chandra\ observations we conducted an extensive
ground-based campaign on \grb\ at radio, optical and near-infrared
wavelengths using the VLA, the \mbox{40-inch} Swope and
\mbox{100-inch} Du~Pont telescopes at Las Campanas Observatory, and
the \mbox{8.2-m} Subaru Telescope on Mauna Kea.  A complete list of
these observations is given in Table~\ref{tab:fluxes}, along with
upper limits on the flux of the afterglow at these epochs and
measurements of the host galaxy brightness.

\begin{figure}
\centerline{~\psfig{file=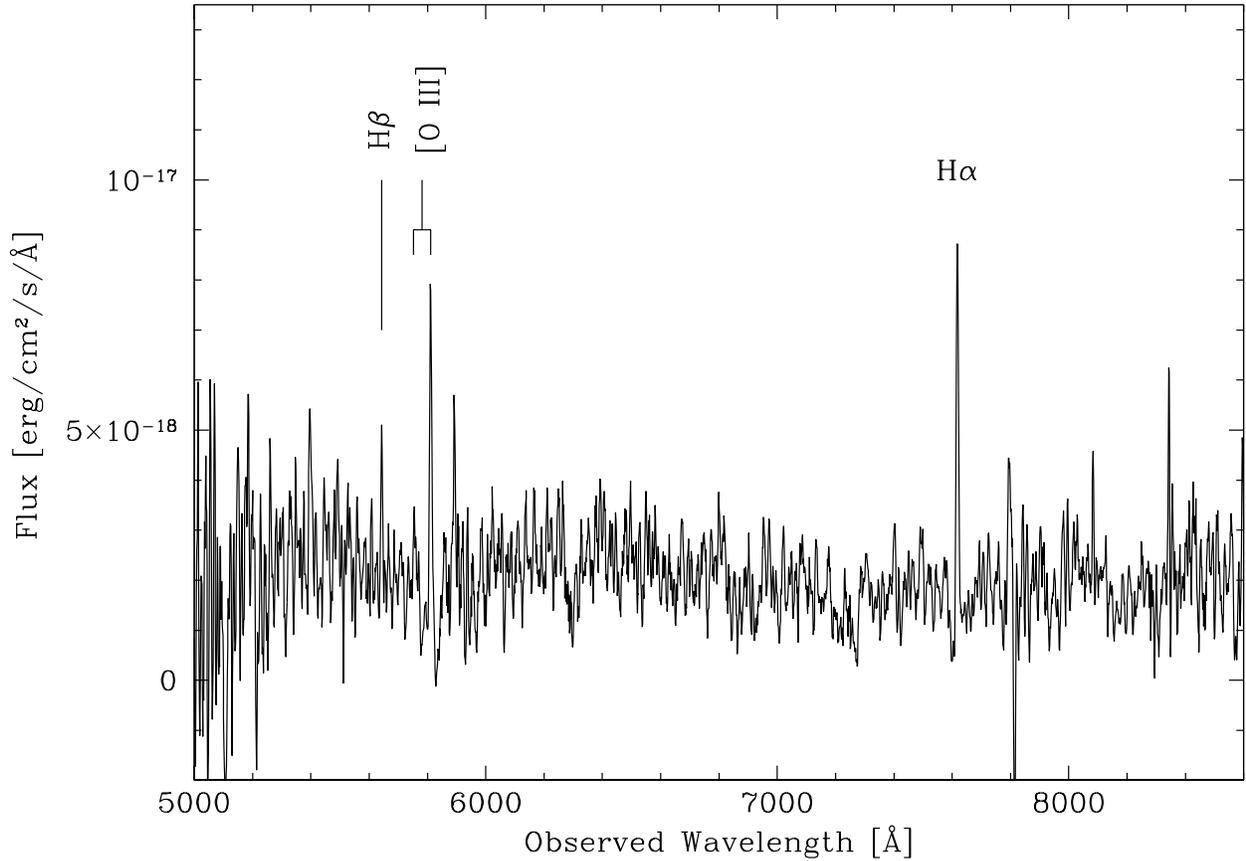,width=7.2in,angle=-90}~}
\caption[]{\bf Spectrum of the host galaxy of \grb.}
\label{fig:spectrum}
\end{figure}

Our \chandra\ afterglow candidate was found to be coincident with a
point-like optical source\cite{ffc+05}, distinct from the candidate
host galaxy, which faded in a manner consistent with the optical
afterglows of long-duration GRBs\cite{hwf+05}. We undertook
spectroscopy of the candidate host galaxy with the Gemini Multi-Object
Spectrograph on the Gemini North telescope, and find it to be a
star-forming galaxy at redshift $z = 0.160$
(Fig.~\ref{fig:spectrum}).

We also triggered a sequence of \hst\ observations with the Advanced
Camera for Surveys\cite{sjb+05} (ACS). Within the \chandra\ error
circle we find a single bright, fading, point-like source, the
unambiguous optical afterglow of \grb; our \hst\ photometry is
presented in Table~\ref{tab:fluxes}.  Expressing the afterglow decay
as a power-law (flux $\propto t^\alpha$) between each epoch, we find
power-law indices of $\alpha_{12}=-1.25\pm 0.09$ between epochs 1 and
2, $\alpha_{23}=-2.83\pm 0.39$ between epochs 2 and 3, and
$\alpha_{34}<-0.43$ between epochs 3 and 4.  Our observation of a
break in the decay is definitive; the \hst\ data are inconsistent with
a single power-law decay at the 3.7-$\sigma$ level.

As can be seen from Figure~\ref{fig:HST}, the optical afterglow of
\grb\ is superposed on the outskirts of the $z=0.16$ candidate host
galaxy.  This precise localization, the first subarcsecond
localization for any short-hard burst, unambiguously associates \grb\
with the $z=0.16$ galaxy.  Thus we show here, for the first time, that
some SHBs arise in low-redshift star-forming galaxies.

The morphology of the host galaxy is irregular, typical of
star-forming galaxies. We fit the radial light profile and find that
it is well described by an exponential disk with scale length
$r_e=0.75\arcsec$.  The afterglow is situated 1.38~arcsec or 1.8$r_e$
from the brightest central region of the galaxy, corresponding to a
physical offset of 3.8\,kpc.  From the detected emission lines we
derive a star formation rate of 0.2\,\Msunyr\ (a lower limit after
allowing for extinction). By comparison, long-duration GRBs are found
exclusively in late-type (star-forming) host galaxies,\cite{bkd02} and
with a somewhat smaller median offset\cite{bkd02} of 1.0$r_e$.


\section{Burst and afterglow energetics}

At a redshift of $z=0.16$, the isotropic-equivalent energy release in
$\gamma$-rays\cite{vlr+05} over the 25--2000\,keV band is
$E_{\gamma,{\rm iso}} = 6.9^{+1.0}_{-0.5}\times 10^{49}$\,erg and the
peak luminosity is $L_p=(1.1\pm 0.5)\times 10^{51}$\,\ergsec\ (here
and throughout this paper, we adopt a flat cosmology with $H_0 =
71$\,\kmsecmpc, $\Omega_{\rm M} = 0.27$ and $\Omega_\Lambda=0.73$).
The burst of $\gamma$-rays is followed by an \xray\ flare, detected by
\hete\ from 25\,s to 130\,s after the burst\cite{vlr+05}. The fluence
of this \xray\ flare is twice as much as that of the $\gamma$-ray
burst itself. Thus, the total isotropic energy release in the first
few hundred seconds is $E_{\rm iso}\sim 10^{50}$\,erg, two orders of
magnitude smaller than that seen in long-duration bursts\cite{fks+01}.

\begin{figure}
\centerline{\psfig{file=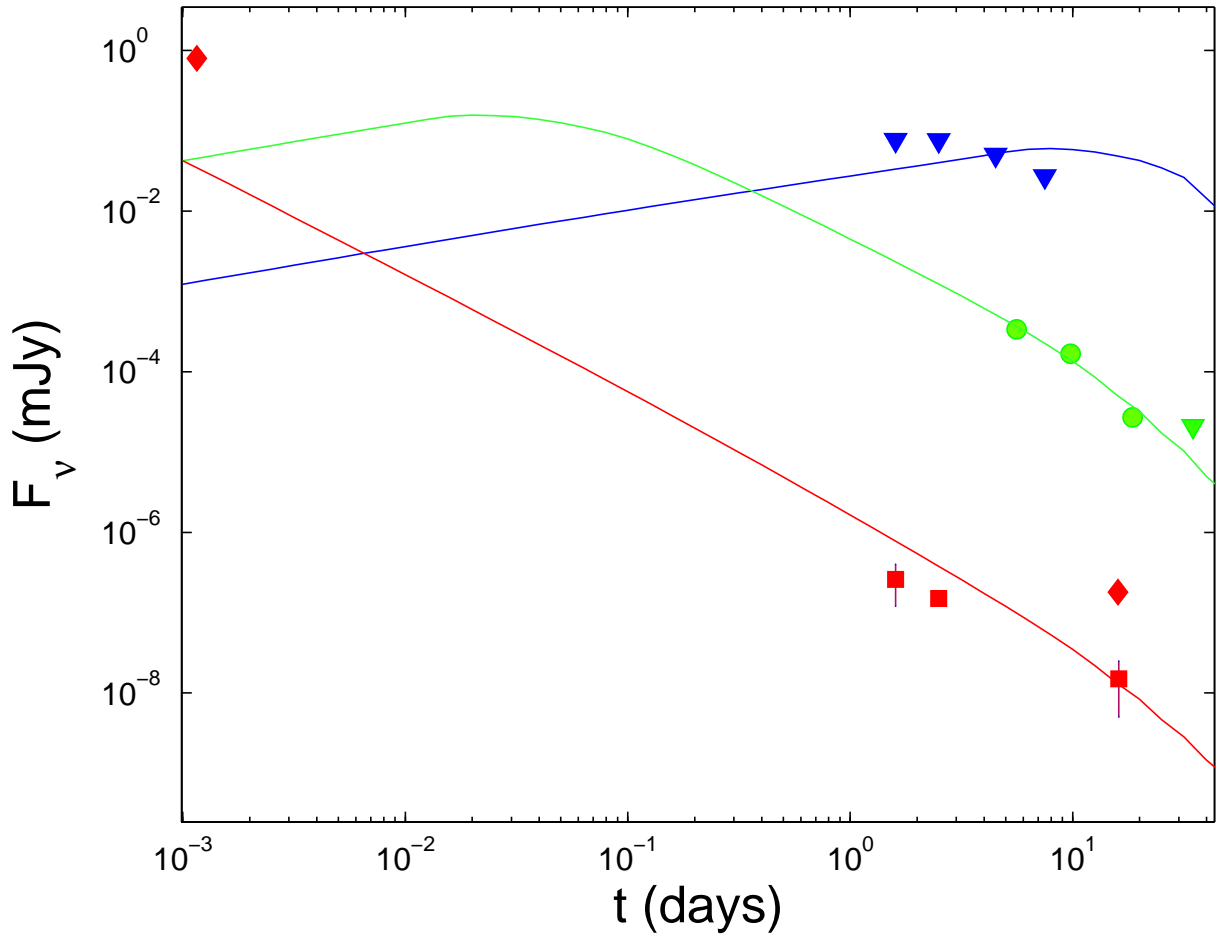,width=7.0in}}
\caption[]{\bf Observations of the \grb\ afterglow and illustrative
  models.}
\label{fig:fnu}
\end{figure}

Figure~\ref{fig:fnu} presents our observations of the \grb\ afterglow.
The steepening power-law decay seen in our \hst\ observations, a
familiar feature of long-duration GRB light curves, is usually
explained as arising from the collimated or jet-like nature of the
ejecta.\cite{rhoads99} This is the first observation of such a
light-curve ``jet break'' for a short-hard burst, although the steep
($\alpha\approx -2$) decay of the GRB\,050724 afterglow is
suggestive.\cite{bpc+05} The epoch of steepening, $t_b\approx 10$\,d,
can be related to the opening angle of the jet, $\theta_j$ (in
radians), as follows\cite{fks+01}:
\begin{equation}
\theta_j = 0.076 \Biggl(\frac{t_b}{\mbox{1 day}}\Biggr)^{3/8}
\Biggl(\frac{1+z}{2}\Biggr)^{-3/8}
\Biggl(\frac{\eta_\gamma}{0.2}\Biggr)^{1/8}
\Biggl(\frac{n}{\rm 1\,cm^{-3}}\Biggr)^{1/8}
\Biggl[ \frac{E_{\gamma,{\rm iso}}}{10^{53}\,{\rm erg}}\Biggr]^{-1/8}\,
{\rm radian};
\end{equation}
here $n$ is the circumburst particle density and $\eta_\gamma$ is the
ratio of the radiated energy to the energy in relativistic ejecta.
Setting $E_{\gamma,{\rm iso}}=10^{50}$\,erg and
$n=10^{-2}$\,\percmcube\ we obtain $\theta_j=0.25$ and a beaming
fraction, $f_b=1-\cos(\theta)\approx 0.03$.  With this value of $f_b$,
the energy released in the first few hundred seconds is $3\times
10^{48}$\,erg -- two orders of magnitude less than that inferred for
long-duration GRBs.\cite{fks+01}

We interpret the fading optical and \xray\ emission to be the
afterglow of GRB\,050709. The afterglow phenomenon has been studied
both observationally and theoretically in the context of long-duration
GRBs\cite{zm04}. The emission arises from ambient material shocked by
the relativistic ejecta.  The broadband spectral index between the
optical and \xray\ bands at late times (\chandra\ and \hst\ 
observations) is $\beta_{\rm ox}=-1.1$; since the temporal power-law
index is $\alpha_{\rm o} = -1.25$ in the optical prior to the jet
break, the electron power-law index is derived to be p=2.7, and we
find that the synchrotron cooling frequency is between the optical and
\xray\ bands at this time.

The isotropic-equivalent \xray\ luminosity $L_{\rm X,iso}$ of the
afterglow at a fixed time after the burst serves as a useful
proxy\cite{kumar00} for the kinetic energy of the ejecta in the
afterglow phase. Extrapolating back from the first \chandra\ flux to a
fiducial time of 10~hr post-burst, assuming a $t^{-1.3}$ decay
(Fig.~\ref{fig:fnu}), we find $L_{\rm X,iso}\sim 2\times
10^{42}$\,\ergsec, at least three orders of magnitude smaller than is
typical for the afterglows of long-duration GRBs\cite{bkf03}.


\section{Limits on an associated supernova}

At $z=0.16$, the distance modulus of \grb\ is $(m-M) = 39.4$\,mag.
Our \hst\ data (Fig.~\ref{fig:HST}) are consistent with a pure
afterglow evolution, and show no evidence for a supernova.  Since we
detect the afterglow in optical light, we are in the unique position
(for SHBs) of being able to exclude any role for extinction in
suppressing the light of an associated supernova.  We estimate our
sensitivity limit for an SN bump as the faintest magnitude we observe,
$m_{\rm SN} > 27.5$\,mag.  Converting to absolute magnitudes and
applying a $k$-correction (from F814W AB to Vega magnitudes) of
$-$0.12\,mag, we estimate $M_R > -12.0$\,mag for any associated SN at
an age of 16\,d.  This limit is fainter than any supernova yet
detected in the nearby Universe.

Numerical studies of SHBs\cite{jr02} predict a modest associated
nova-type event, much fainter than the average supernova. Quantitative
predictions for the ejecta mass, speed, and composition span an
extremely broad range, however, so that the situation is ripe for
observational inputs.

In the absence of heat input after the explosion, adiabatic expansion
of any sub-relativistic ejecta results in very rapid
fading\cite{lp98,k05}. Continued heat input might by provided by the
decay of radioactive nuclei (including nickel)\cite{lp98}, decay
of free neutrons\cite{k05}, or a long-lived central engine\cite{k05}.
Any such heat input will be reprocessed to lower energies -- mainly
via electron Thompson scattering -- provided the ejecta are dense
enough.  The high sensitivity of current facilities makes optical
wavelengths the band of choice for these searches.

Our observations constrain the kinetic energy of slow ejecta in these
models\cite{k05} to be less than $10^{49}$\,erg, provided that the
ejecta velocity $v\simlt 0.02 c$.  The current limit arises entirely
from the first-epoch \hst\ data. Sensitive optical data taken at
earlier times would have provided stronger constraints and for higher
ejecta speeds.  

An alternative source of heat could be luminosity from a long-lived
central engine.  The \hst\ observations constrain this luminosity to
be $L\simlt 10^{41}$\,\ergsec. We remark that the X-ray flare in the
second epoch \chandra\ data has a similar luminosity.  


\section{Properties of short-hard bursts}

The offsets of \grb\ and GBB\,050724 from their proposed host galaxies
are small enough that their associations can be considered secure.
These two associations, in turn, strengthen the case for
identification of GRB\,050509B (localized to 9.3-arcsec
radius\cite{gbb+05}) with the redshift $z=0.225$ galaxy\cite{bpp+05}
and galaxy cluster\cite{gbb+05} that have been proposed to host this
burst. We can thus set the physical scale for the energetics of all
four bursts.  Table~\ref{tab:global} presents the properties of these
SHBs, along with some properties of their host galaxies.
Figure~\ref{fig:histo} places these bursts -- the only known SHB
afterglows -- in the context of the set of long-duration GRBs with
known redshifts.

In Table~\ref{tab:global} we display the peak luminosities of the four
SHBs, extrapolated to the full BATSE band for comparison with results
from that experiment.  All four values are approximately
$L_{\gamma,{\rm peak}}\sim 10^{50}$\,\ergsec.  The similarity of the
redshifts and peak luminosities of the SHBs suggests that they arise
from a single source population.

Next, from Figure~\ref{fig:histo} we see, relative to the long
duration GRBs, SHBs are located at lower redshift, emit less energy
in $\gamma$-rays, and possess a less-energetic afterglow.  This
behaviour is broadly consistent with conclusions from earlier
statistical studies\cite{totani99,schmidt01,bbh+03,gp05}.

A closer distance scale for the SHBs is consistent with the value of
$\langle V/V_{\max}\rangle \approx 0.4$ for BATSE
SHBs,\cite{kc96,gp05} which is significantly higher than the BATSE
value for long-duration GRBs. In particular, the BATSE SHB $\langle
V/V_{\max}\rangle$ value is consistent with a spatially-uniform
distribution of standard candles out to $z_{\max}\simeq
0.4$\cite{piran96} and with peak luminosities of $L_{\gamma,{\rm
peak}}\approx 10^{50}$\,\ergsec\ as observed for these four bursts.

In Table~\ref{tab:global} we also summarize selected properties of the
four SHB host galaxies.  The proposed host of GRB\,050509B is a large
elliptical galaxy at redshift $z=0.2248$,\cite{bpp+05} with luminosity
$L\approx 1.5L_*$ (ref.\ \pcite{GCN.3418}).  The galaxy has little
ongoing star formation, $< 0.1$\,\Msunyr\ (ref.\ \pcite{bpp+05}).  The
elliptical host galaxy of GRB\,050724 (which has also been localized
to sub-arcsecond precision) is a luminous ($L=1.6L_*$) elliptical
galaxy with star formation rate $< 0.02$\,\Msunyr (ref.\
\pcite{bpc+05}).  

In contrast, the morphology and spectrum of the host galaxy of
GRB\,050709 (Figure~\ref{fig:HST}) indicates that the host is an
irregular, late-type galaxy with a significant star formation rate,
$\sim 0.2$\,\Msunyr, and a luminosity much smaller than the other SHB
hosts, $L\approx 0.10 L_*$; thus the galaxy is forming roughly as many
stars per unit stellar mass as the Milky~Way.

The association of SHBs with both star-forming galaxies and early-type
ellipticals is reminiscent of the diversity of host galaxies of
type~Ia supernovae (SNe).  As with the type~Ia SNe, this dichotomy may
indicate a class of progenitors with an extremely wide range of
lifetimes between formation and explosion, with some systems living
for many gigayears. However, even though type~Ia SNe occur in
elliptical galaxies, the rate of such events is higher in late-type,
star-forming galaxies, and the majority of type~Ia events in the local
Universe occur in late-type blue galaxies\cite{mvp+05}. The trend
emerging for the SHBs, in which the ``majority'' of events (here
perhaps, 3 of 4) occur in elliptical galaxies, could indicate that the
progenitor systems are even longer-lived than those of Ia supernovae.


\section{The nature of the short-hard bursts}

\begin{figure}
\centerline{~\psfig{file=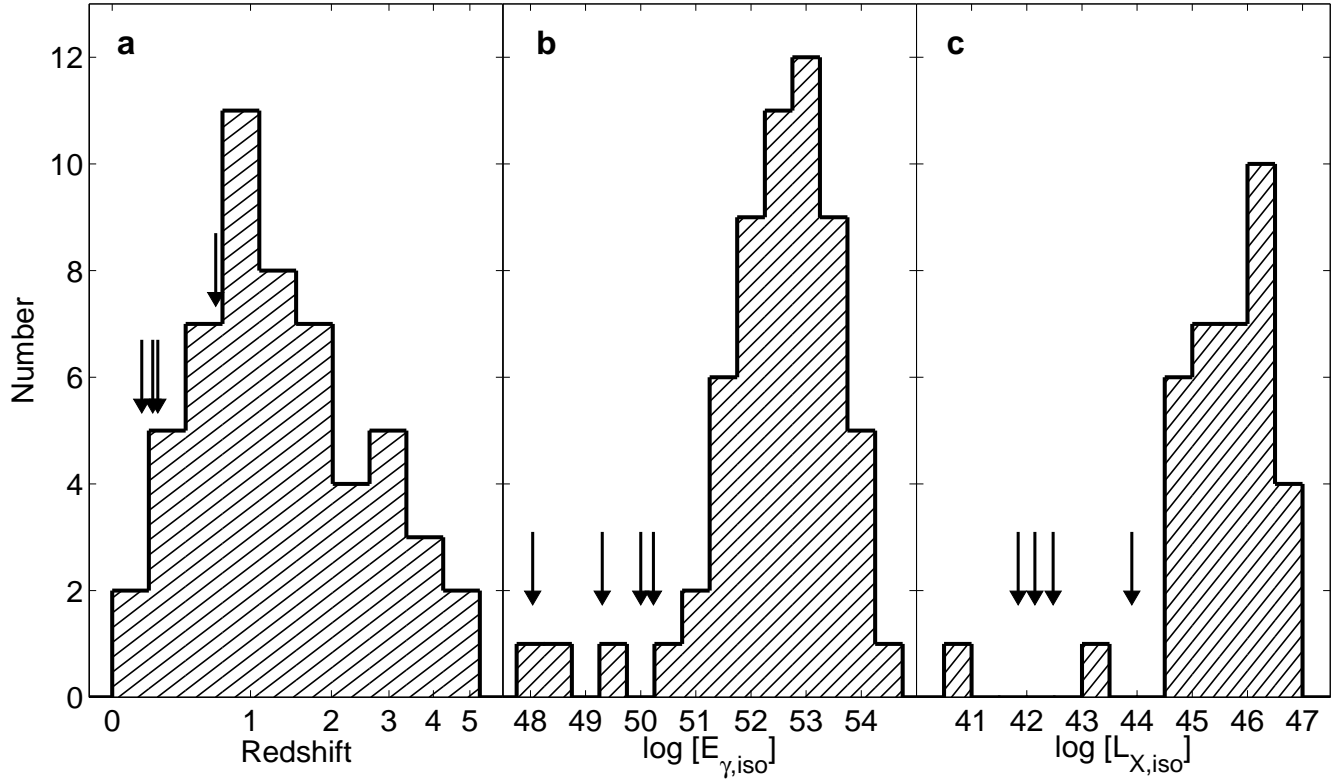,width=7.0in,angle=0}~}
\caption[]{\bf Physical properties of the afterglows of long-duration GRBs
    (histograms) and SHBs (arrows).}
\label{fig:histo}
\end{figure}

Our observations of \grb\ support the view -- until recently founded
only on the basis of their prompt $\gamma$-ray emissions -- that the
SHBs are a different population from the long-duration GRBs
(Fig.~\ref{fig:histo}).  They are lower-energy explosions, with a
correspondingly less-energetic relativistic blastwave, and they are
found at significantly closer distances.  At the same time, the
similarity of the SHB redshifts and peak luminosities to one other
(Table~\ref{tab:global}) strongly suggests a common origin for these
events.

We find that SHBs are distinctly weaker explosions than long-duration
GRBs. The lower energies that we infer from afterglow observations of
GRB\,050709 and GRB\,050724 are consistent with the merger of a
compact object binary, as seen in numerical simulations.  Moreover,
the two best-studied events show strong collimation; thus the true
rate of SHBs is 30 to 100 times the observed rate.

The long-duration GRBs have been definitively associated with the
deaths of massive stars\cite{kfw+98,gvv+98,hsm+03,smg+03}. Two
properties of the known SHBs argue against such an association: first,
the fact that GRB\,050724 occurred in an elliptical galaxy without
active star formation,\cite{bpc+05} and that GRB\,050509B likely
occurred in an elliptical as well;\cite{bpp+05} and second, that
GRB\,050509B\cite{kfc+05,hsg+05} and \grb\ lack associated supernovae,
a common feature of $z<1$ long-duration GRBs, to very deep limits.
Separately, the presence of SHBs among old stellar populations in
elliptical galaxies argues against a magnetar origin, or a form of
short-lived compact binary.

The locations of the SHBs with respect to their host galaxies are
compatible with the kicks delivered to neutron star binary systems at
birth\cite{npp92}. Their occurrence in both star-forming (late-type)
and non-star-forming (early-type) galaxies suggests that there may be
a substantial range of lifetimes for the progenitor systems, perhaps
extending to many gigayears.

In all respects, the emerging picture of SHB properties is consistent
with an origin in the coalescence events of neutron star-neutron star
or neutron star-black hole binary systems.  The stage is now set for
detailed studies of these exotic cosmic explosions, the most exciting
of which would be the detection of their associated bursts of
gravitational waves.



\bibliographystyle{nature-pap}
\bibliography{journals,refshb}


\begin{acknowledge}
  Our GRB research is supported in part by funds from NSF, NASA, the
  Australian Research Council, and the Ministry of Education, Science,
  Culture, Sports, and Technology in Japan. The VLA is operated by the
  National Radio Astronomy Observatory, a facility of the NSF operated
  under cooperative agreement by Associated Universities, Inc.  The
  Gemini Observatory is operated by the Association of Universities
  for Research in Astronomy (AURA), Inc., under a cooperative
  agreement with the NSF on behalf of the Gemini partnership. This
  work is based in part on data collected at Subaru Telescope, which
  is operated by the National Astronomical Observatory of Japan.

\end{acknowledge}


\newpage


\begin{table}
\begin{center}
\begin{tabular}{lllllll}
\hline
\hline
UT Date     & UT    &$\Delta T$&Facility& Band & Exp & Flux \cr
\hline
\hline
11-Jul-2005 & 04:20 &  1.24\,d & Swope-40 & $i'$ & $3\times 600\,$s & $>20.5$\,mag\cr
11-Jul-2005 & 10:19 &  1.49\,d & Swope-40 & $i'$ & $3\times 600\,$s & .. \cr
13-Jul-2005 & 08:33 &  3.41\,d & Du Pont-100 & $R$ & $3\times 600$\,s & ($21.05\pm 0.15$\,mag)\cr
15-Jul-2005 & 14:06 &  5.6\,d  & Subaru & $K'$ & $270\times 20$\,s  & $22.1\pm 0.7$\,mag\cr
26-Jul-2005 & 13:19 & 16.6\,d  & Subaru & $K'$ & $360\times 20$\,s & ($19.2\pm 0.1$)      \cr
\hline
11-Jul-2005 & 12:14 &  1.6\,d & VLA & 8.46\,GHz & 6660\,s & $<$76.6\,\uJy\cr
12-Jul-2005 & 09:36 &  2.5\,d & VLA & 8.46\,GHz & 1265\,s & $<$75.8\,\uJy\cr
14-Jul-2005 & 11:31 &  4.5\,d & VLA & 8.46\,GHz & 6055\,s & $<$49.4\,\uJy\cr
17-Jul-2005 & 10:48 &  7.5\,d & VLA & 8.46\,GHz & 6490\,s & $<$26.6\,\uJy\cr
\hline
09-Jul-2005 & 22:38 &  100\,s & \hete\ WXM & 5\,keV & 10\,s & $0.80\pm 0.14$\,mJy\cr
\hline
12-Jul-2005 & 11:15 &  2.5\,d & Chandra & 5\,keV & 38.4\,ks & $0.15\pm 0.02$\,nJy\cr
25-Jul-2005 & 21:47 & 16.0\,d & Chandra & 5\,keV &  6.1\,ks & $0.18\pm 0.06$\,nJy\cr
26-Jul-2005 & 00:18 & 16.1\,d & Chandra & 5\,keV & 12.1\,ks & $0.015^{+0.027}_{-0.007}$\,nJy\cr
\hline
11-Jul-2005 & 13:07 &  1.6\,d & Swift & 5\,keV & 15.2\,ks & $0.26\pm 0.14\,$nJy\cr
12-Jul-2005 & 10:13 &  2.4\,d & Swift & 5\,keV &  5.0\,ks & $<$0.2\,nJy\cr
13-Jul-2005 & 02:32 &  3.2\,d & Swift & 5\,keV &  6.4\,ks & $<$0.1\,nJy\cr
14-Jul-2005 & 05:26 &  4.3\,d & Swift & 5\,keV &  2.1\,ks & $<$0.8\,nJy\cr
\hline
15-Jul-2005 & 13:49 &  5.6\,d & HST   & F814W & 6360\,s & $25.08\pm 0.02$\,mag\cr
19-Jul-2005 & 17:11 &  9.8\,d & HST   & F814W & 6360\,s & $25.84\pm 0.05$\,mag\cr
28-Jul-2005 & 13:48 & 18.6\,d & HST   & F814W & 6360\,s & $27.81\pm 0.27$\,mag\cr
13-Aug-2005 & 15:17 & 34.7\,d & HST   & F814W & 6360\,s & $>$28.1\,mag\cr
\hline
\hline
\end{tabular}
\end{center}

\caption[]{\bf Observations of the afterglow of GRB\,050709}
\label{tab:fluxes}
\end{table}
{\small\noindent%
We carried out all observations summarized in this table except those
of \hete\ (from ref.\ \pcite{vlr+05}) and \Swift\ (reported in ref.\
\pcite{Morgan05}).  The \hete\ point represents the average flux in
the ``soft flare'' that occurs 100\,s after the main burst.  We have
made an independent reduction of the \Swift\ XRT data. \\
\noindent{\bf Optical \&\ Near Infrared.}
The optical and near infrared fluxes do not contain a correction for
Galactic extinction which is expected to be rather small in this
direction\cite{sfd98}: $E(B-V) = 0.012$\,mag. Near-infrared
observations were made with the Cool Infrared Spectrograph and Camera
for OH Suppression\cite{mim+02} (CISCO).  No flux is reported for the
\mbox{Swope-40} second epoch since the image was subtracted from the
first epoch in order to search for the afterglow. Host fluxes in the
table are given in parentheses.  \\
\noindent{\bf Radio.}
The VLA data were taken in standard continuum mode with a bandwidth of
$2 \times 50$ MHz centered in the 8.46-GHz band.  We used 3C48 for
flux calibration and phase referencing was performed against
calibrator J2257$-$364.  Data were reduced using standard packages
within the Astronomical Image Processing System (AIPS). Within the
HETE error circle we find a single source with constant flux ($644\pm
24\,\mu$Jy) at coordinates $\alpha$=23:01:32.1$\pm 0.003$,
$\delta=-$38:59:26.8$\pm 0.1$ (J2000).  This source is coincident with
the cataloged extended NVSS source and resolved \chandra\ source.
Upper limits are quoted at 2$\sigma$ or 95.5\% confidence.\\
\noindent{\bf Chandra X-ray Observatory.}
Our basic data reduction procedures are described in the text.  We use
a custom pipeline composed of CIAO 3.2.1 tools to run a full
``wavdetect'' analysis on images constructed from the 0.3--2.0 keV,
2.0--8.0 keV, and 0.3--8.0 keV bands separately, and then merge the
resulting source catalogs.  For spectral fits, photons are extracted
from a 1.5\arcsec-radius aperture, and fits are made using XSPEC 12.0.
The second observation has been divided into ``flaring'' and
``quiescent'' intervals; see text for details. \\
\noindent{\bf Swift X-ray Telescope.}
The photometry was done using a circular region of radius
0.8\,arcmin centered on the \xray\ afterglow position. The
background was estimated from the entire XRT image.  The detection
confidence level from Poisson statistics is 2.3$\sigma$ for the first
observation; the remaining observations provide only upper limits
(quoted at 2$\sigma$ or 95.5\%-confidence).  Photon count limits are
converted to flux densities using our spectral fit to the first-epoch
\chandra\ data; see text for details. \\
\noindent{\bf Hubble Space Telescope.}
Data were obtained with the ACS instrument aboard HST\cite{sjb+05}.
Each epoch consisted of a series of exposures in the F814W (I-band)
filter with a total integration time of 6360\,s.  The images were
processed using Archive ``on-the-fly'' processing, drizzled\cite{fh02}
to the native pixel scale of 0.05 arcsecond, and combined.  We
subtracted the fourth-epoch image from each previous epoch and
performed photometry using a 0.15\arcsec\ (3-pixel) aperture.  The
magnitudes, quoted in the ``AB'' system, are corrected for finite
aperture (0.32\,mag) and imperfect charge transfer efficiency
(0.01\,mag).  Errors are estimated by photometering multiple
background regions in the subtracted images using the same aperture.
The limit on afterglow flux in the fourth epoch is derived by
subtracting point sources of decreasing flux from the image until a
flux deficit at the afterglow position is no longer readily
discernible.}


\newpage


\begin{table}
\begin{center}
\begin{tabular}{lcccc}
\hline
  Property     & GRB\,050509B & GRB\,050709  
               & GRB\,050724  & GRB\,050813 \cr\hline
  Redshift     & 0.225
               & 0.160   
               & 0.258
               & 0.722  \cr

  $T_{90}$     &  $40\pm 4$ ms
               &  $70\pm 10$ ms
               &  $3\pm 1$ s 
               &  $0.6\pm 0.1$ s   \cr

  Fluence ($\ergcmsq$)
               & 9.5$\times 10^{-9}$
               & 2.9$\times 10^{-7}$
               & 6.3$\times 10^{-7}$ 
               & 1.2$\times 10^{-7}$  \cr

  Fluence band & 15--350 keV & 30--400 keV 
               & 15--350 keV & 15--350 keV  \cr

  $E_{\gamma,{\rm iso}}$ (erg)
               & 4.5$\times 10^{48}$   
               & 6.9$\times 10^{49}$ 
               & 4.0$\times 10^{50}$
               & 6.5$\times 10^{50}$   \cr

  $L_{\gamma,{\rm peak}}$ (erg\,s$^{-1}$)
                & $1.4\times 10^{50}$
                & $1.1\times 10^{51}$
                & $1.7\times 10^{50}$ 
                & $1.9\times 10^{51}$   \cr

  $L_{\rm X}$ (\ergsec)
               & $<$7$\times 10^{41}$
               & 3$\times 10^{42}$
               & 8$\times 10^{43}$ 
               & 9$\times 10^{43}$  \cr

  $f_b$        & -- & 0.03 & 0.01 & -- \cr

  $E_{\gamma}$ (erg) & $<4.5\times 10^{48}$
                     & $2.1\times 10^{48}$
                     & $4.0\times10^{48}$ 
                     & $<6.5\times 10^{50}$  \cr

  Host $L/L_*$ & 1.5 & 0.10 & 1.6 & -- \cr

  Host SFR (\Msunyr)
               & $<$0.1 & 0.2 & $<$0.03 & -- \cr

  Offset (kpc)
               & $44^{+12}_{-23}$ & 3.8 & 2.6 & -- \cr

  Offset ($r/r_e$)
               & $13^{+3}_{-7}$   & 1.8 & 0.4 & -- \cr

  SN Limit, $M_R$ (mag)
               & $> -13.0$ & $> -12.0$ & -- & -- \cr\hline

\end{tabular}
\end{center}
\caption[]{\bf Physical properties of short-hard bursts and their host 
  galaxies}
\label{tab:global}
\end{table}
{\small\noindent%
  In order, we show: the burst redshift; the duration of the
  90\%-inclusive interval of high-energy emission ($T_{90}$); the
  measured burst fluence and corresponding energy bandpass; the peak
  burst luminosity; the isotropic-equivalent $\gamma$-ray burst
  energy; the isotropic-equivalent luminosity in \xray\ at 10\,h
  post-burst; the beaming fraction, calculated from the jet
  collimation angle $\theta_j$ as $f_b = 1-\cos(\theta_j)$; the
  $\gamma$ ray energy release corrected for beaming fraction, $f_b
  E_{\gamma,{\rm iso}}$; the host galaxy luminosity; the host
  star-formation rate; the burst offset from its host, in physical
  units and referred to the scale length of the host galaxy's light
  profile; and the absolute magnitude limit on any associated
  supernova.  Blank entries (``--'') are unconstrained by the data at
  present; values without uncertainties are known to roughly 20\%
  precision.  References: GRB\,050509B, refs.\ \pcite{gbb+05},
  \pcite{bpp+05} and \pcite{kfc+05}; GRB\,050709, ref.\ \pcite{vlr+05}
  and this work; GRB\,050724, refs.\ \pcite{Covino05} and
  \pcite{bpc+05}; GRB\,050813, refs.\ \pcite{fr+05} and \pcite{gb+05}.
  The values of $E_{\gamma,{\rm iso}}$ and $L_{\gamma,{\rm peak}}$
  have been increased by a factor of 4, representing a mean correction
  from the BATSE sample for converting these observed fluences to the
  25--2000 keV band.}

\clearpage



{\small\noindent%
(a) The Chandra (0.3--8.0 keV) image of the field from our observation
of 2005 July 12.5 UT.  The large circle is the \hete\ localization
region, 81\,arcsec in radius; north is up, east is to the left, and
the scale of the image is indicated.  A red ellipse indicates the
faint \xray\ source which we identify with an NVSS catalog object; the
bright point source in the boxed region is the afterglow of \grb.  (b)
Close-up of the region surrounding the \xray\ afterglow, in a
coaddition of all our \hst\ data; the red circle is the \chandra\
localization region, 0.5\,arcsec in radius.  A point source is visible
within this region; the source is observed to fade over the course of
our \hst\ observations, and we identify it as the optical afterglow of
\grb.  The irregular galaxy to its west is the proposed $z=0.16$ host
galaxy.  We use the GALFIT software\cite{phi+02} to fit the radial
surface brightness distribution of the host galaxy, with the Sersic
concentration parameter, $n$, and the effective radius, $r_e$, as free
parameters.  We find a best-fit solution ($\chi2\approx 3$ per degree
of freedom) with $n=1.1$ and $r_e=0.75''$.  At $z=0.16$ the
afterglow's 1.38-arcsec offset from the brightest region of this
galaxy corresponds to 3.8~kpc in projection, or 1.8$r_e$.}

\clearpage


{\small\noindent%
These observations were taken with GMOS on the Gemini North telescope
under poor seeing and non-photometric conditions.  We obtained a total
integration of $3,303\,$s in four exposures before closing due to the
onset of twilight.  The spectra were processed with the GMOS data
reduction package in IRAF, sky-subtracted, combined and extracted, and
smoothed with a 7\AA\ boxcar. Flux calibration was performed using an
observation of Hiltner 600 observed with a similar instrument setup in
a previous lunation; hence the flux scale shown is indicative rather
than absolute.  Three emission lines are readily apparent (H$\alpha$,
[OIII]\,$\lambda\,5007$ and H$\beta$) yielding a mean redshift of $z =
0.160 \pm 0.001$.  The observed H$\alpha$/H$\beta$ flux ratio of 3.1
indicates that the global galaxy extinction is small ($A_V =
0.1$~mag).  }

\clearpage


{\small\noindent%
  The X-ray (red), optical (green) and radio (blue) data taken from
  Table~\protect\ref{tab:fluxes}. These multi-wavelength observations
  can be marginally accommodated within a standard external-shock
  afterglow model. The best fit parameters are $E=5 \times 10^{48}$
  erg, $\epsilon_e=\epsilon_B=1/3$, $n=0.01$, $p=2.5$ and
  $\theta_j=0.35$.  However even this best fit predicts a radio flux
  which is slightly higher (a factor of 2) than our upper limits and
  requires an uncommonly high value for $\epsilon_B$. The latter is a
  concern because we do not expect the microphysics of the afterglow
  to differ between the short- and long-duration burst populations,
  although one cannot exclude this possibility. The early \hete\ X-ray
  observation and the late \chandra\ \xray\ flare cannot be explained
  by the standard external shock afterglow model, both are too
  bright. The early point can be a flare resulting from an energy
  injection to the external shock or a long lasting activity of the
  source, as seen recently by the Swift/XRT in a long
  burst\cite{brf+05}.  The late flare however is a unique phenomenon
  to short bursts that was not observed so far in any long burst.  The
  duration of the flare ($\sim 0.1$ day) corresponds to an emitting
  region at a size of $10^{14}$cm 16 days after the burst. At this
  time the external shock is mildly relativistic and its width is
  $\sim 10^{16}$ cm. The (isotropic equivalent) energy emitted in this
  flare is $\sim 10^{45}$ erg in the \xray\ alone. These different
  length scales, together with brightness of the flare, exclude the
  possibility that the flare is produced in a simple external shock
  and indicates a process that was not observed so far in long bursts.
  Most probably, this process involves an activity of the source at $t
  \sim 16$~days which is $10^7$ times the duration of the
  burst. Combined with the unique \xray\ light curve of the short-hard
  burst GRB\,050724 (showing an early very steep decay and a later
  brightening) these observations suggest that one should keep an open
  mind for the possibility that SHB afterglows are a drastically
  different phenomenon than afterglows of long GRBs.}

\clearpage


{\small\noindent%
a--c, For GRBs with known redshifts, their distribution in redshift
(a), isotropic-equivalent gamma-ray luminosity ($\log[E_{\gamma,{\rm
iso}}]$, b), and isotropic-equivalent \xray\ luminosity at 10~hours
after the burst ($\log[L_{\rm X,iso}]$, c).  Arrows indicate the
values of these properties for the four SHBs with afterglow
detections, GRB\,050509B, GRB\,050709, GRB\,050724, and GRB\,050813
(as they are ordered consistently, from lowest to highest, in each
panel). The contrast between the physical properties of the two burst
populations is dramatic.  GRB properties are from refs\ \pcite{bkf03},
\pcite{bkp+03}, \pcite{fb05}, and \pcite{bkf+05}.  See
Table~\protect\ref{tab:global} for SHB references.}

\clearpage



\end{document}